\documentclass[preprint2]{aastex}

\newcommand{\etal}{et~al.}

\newcommand{\Fefs}{$^{56}$Fe}
\newcommand{\Cofs}{$^{56}$Co}
\newcommand{\Nifs}{$^{56}$Ni}
\newcommand{\feii}{[Fe {\sc ii}]}
\newcommand{\coiii}{[Co {\sc iii}]}

\newcommand{\oi}{[O {\sc i}]}

\def\gsim{\mathrel{\rlap{\lower 4pt \hbox{\hskip 1pt $\sim$}}\raise 1pt
\hbox {$>$}}}
\def\lsim{\mathrel{\rlap{\lower 4pt \hbox{\hskip 1pt $\sim$}}\raise 1pt
\hbox {$<$}}}

\shorttitle{Asymmetric Explosion of SNe Ia}
\shortauthors{Motohara \etal}

\begin{document}

\title{Asymmetric Explosion of Type I\lowercase{a} Supernovae as Seen from Near 
Infrared Observations\footnotemark[1]}

\author{
 Kentaro Motohara\altaffilmark{2}, 
 Keiichi Maeda\altaffilmark{3},
 Christopher L. Gerardy\altaffilmark{4}, 
 Ken'ichi Nomoto\altaffilmark{5},
 Masaomi Tanaka\altaffilmark{5},
 Nozomu Tominaga\altaffilmark{5},
 Takuya Ohkubo\altaffilmark{5},
 Paolo A. Mazzali \altaffilmark{6,7,5},
 Robert A. Fesen\altaffilmark{8},
 Peter H\"oflich\altaffilmark{9}, and 
 J. Craig Wheeler\altaffilmark{9}
}

\altaffiltext{1}{Based on data collected at Subaru Telescope, which is operated by the National Astronomical Observatory of Japan.}
\altaffiltext{2}{Institute of Astronomy, University of Tokyo, Mitaka, Tokyo 181-0015, Japan; kmotohara@ioa.s.u-tokyo.ac.jp}
\altaffiltext{3}{Department of Earth Science and Astronomy, Graduate School of Arts and Science, University of Tokyo, Meguro-ku, Tokyo 153-8902, Japan}
\altaffiltext{4}{Astrophysics Group, Imperial College, Blackett Laboratory, Prince Consort Road, London SW7 2BZ, UK}
\altaffiltext{5}{Department of Astronomy, University of Tokyo, Bunkyo-ku, Tokyo 113-0033, Japan}
\altaffiltext{6}{Max-Planck-Institut f\"ur Astrophysik, Karl-Schwarzschild-Stra{\ss}e 1, 85741 Garching, Germany}
\altaffiltext{7}{Instituto Nazionale di Astrofisica (INAF)-Osservatorio Astronomico di Trieste, Via Tiepolo 11, I-34131 Trieste, Italy}
\altaffiltext{8}{Department of Physics and Astronomy, 6127 Wilder Laboratory, Dartmouth College, Hanover, NH 03755}
\altaffiltext{9}{McDonald Observatory, University of Texas, Austin, TX 78712}

\begin{abstract}
We present near-infrared spectra of late phase ($>200^{\rm d}$) Type Ia supernovae (SNe Ia) 
taken at the Subaru telescope.
The \feii\ line of SN 2003hv shows a clear flat-topped feature, while
that of SN 2005W show less prominent flatness.
In addition, a large shift in their line center, varying from 
$-3000$ to $1000$ (km s$^{-1}$) with respect to the host galaxies, is seen.
Such a shift suggests the occurrence of an off-center, non-spherical explosion 
in the central region,
and provides important, new constraints on the explosion models of SNe Ia.
\end{abstract}

\keywords{infrared: stars --- supernovae: general --- supernovae: individual (SN 2003du, SN 2003hv, SN 2005W)
\begin{center}
{\bf Accepted for publication in the Astrophysical Journal (Letters)}
\end{center}
}

\section{Introduction and Summary}
The brightness and approximate uniformity of 
Type Ia supernovae (SNe Ia)
enable us to use them as reliable and distant 
standard candles, and 
provided evidence for the accelerating universe
\citep{riess98, perlmutter99}.
For precision cosmology, it is critical to 
understand the origin of diversity, as well as 
the explosion mechanism of SNe Ia 
that is still open to debate 
\citep[see, e.g., ][for a review]{hillebrandt00}.
Possible sources of diversity include 
the asymmetry of the explosion \citep{wang03} due to 
the rotation of the progenitor white dwarfs (WDs) 
\citep{piersanti03, saio04, uenishi03, yoon03}
and/or 
turbulent behavior of the deflagration flame \citep[e.g., ][]{gamezo03,ropke06}.
Therefore, it is important to observationally investigate
the distribution of the synthesized elements and the kinematical structure of the ejecta 
in order to constrain the explosion models. 

In this respect, late-phase ($\sim 1$yr since the explosion) spectroscopy 
at near-infrared (NIR) wavelength provides important diagnostics.
Because the ejecta become optically thin in late phases, 
spectroscopy provides an unbiased, 
direct view of the innermost regions.
Well isolated \feii\ emission lines at 1.257 and $1.644\ \micron$ 
enable us to trace the distribution of the most important isotopes 
synthesized in SNe Ia, i.e., $^{56}$Ni (which decays into $^{56}$Co and then $^{56}$Fe) 
and other iron isotopes such as $^{54}$Fe. 
NIR observations of the type Ia SN~1991T show a strongly peaked \feii\
$1.644\ \micron$ line which appears symmetric with respect to the rest-frame
of the host galaxy \citep{spyromilio92, bowers97}.
The NIR \feii\ line profile of SN~1998bu observed by \citet{spyromilio04}
appears to have a somewhat less centrally-peaked emission profile.

This approach was highlighted by the discovery of 
a flat-topped \feii\ $1.644\ \micron$ emission line in SN 2003du  
\citep{hoeflich04}. 
        Such a line profile requires a central hole in the kinematic
        distribution of the radioactive ejecta.  This sort of hollow
        radioactivity distribution is predicted in 1D explosion models
        \citep[e.g.,][for a review]{nomoto94} in
        which the innermost regions are burned under high densities,
        $\rho > 10^{9} \rm g\,cm^{-3}$,
        and electron capture produces stable isotopes of
        $^{58}$Ni, $^{54}$Fe, and $^{56}$Fe, 
        rather than radioactive $^{56}$Ni.  However, such a distribution is
        contrary to the predictions of state-of-the-art 3D deflagration
        simulations which predict large-scale turbulent mixing in the inner
        layers, and thus do not confine $^{56}$Ni outside the central core. 
The \citet{hoeflich04} NIR spectrum of SN~2003du also indicate that the \feii\ 1.257, $1.644\ \micron$ 
emission lines show a blueshift of $500 - 1,000$ km s$^{-1}$ 
with respect to the host galaxy. 

However, since the observed sample size of 
late-phase NIR spectra is so small, it is not clear whether
the flat-topped and blue-shifted lines are general properties of SNe Ia.
Such observations are extremely difficult because
even relatively nearby and bright SNe Ia 
are faint in NIR at late phases (e.g., $H = 20 \sim 21$ for SN 2003du 
at $\sim$ 1 year after the explosion). 
We therefore have conducted NIR observations of SNe Ia using the Subaru with 
OH airglow suppressor (Table 1) 
and obtained 2 more late-phase $ 1- 2 \mu$m spectra of SN 2003hv and 2005W.

Detailed explanation and discussion of the spectra will 
be presented elsewhere (Motohara et al., in prep.).
We briefly report two interesting results in this letter.
(1) The \feii\ lines of SN~2003hv also show a clear flat-topped feature.
(2) The \feii\ lines of SN~2005W show less prominent flatness.
(3) A large velocity shift of the line center, varying from 
$-3000$ to $1000$ (km s$^{-1}$) with respect to the host galaxies, exists. 

Considering the uniformity in optical bands around 
the peak luminosity, the existence of 
such inhomogeneity and asymmetry in SNe Ia is surprising, 
and provides important, new constraints on the explosion models.

\section{Observations}

\subsection{SN 2003du}
SN 2003du was discovered by LOTOSS on 2003 Apr. 22.4 in UGC 9391 at about 15.9mag \citep{schwartz03}.
It was confirmed to be a SN Ia by \citet{kotak03} who reported that the optical spectrum resembles that of SN 2002bo 
about 2 weeks before maximum, and reached maximum light in the $B$-band 
($t_{B \rm max}=0^{\rm d}$) on 2003 May 6.3 UT (JD $2452766.3 \pm 0.5$) \citep{anupama05}.

As reported \citet{hoeflich04},
$JH$-band spectroscopy of SN 2003du 
was carried out using OHS/CISCO \citep{iwamuro01,motohara02} at Subaru.
For this work, we re-reduced the data taken on 2004 Feb 27.5 ($+297^{\rm d}$),
which were processed in a standard procedure of flat fielding, sky subtraction, 
bad-pixel correction, and residual sky subtraction. 
The flux was scaled using the $H$-band photometry that was taken on
2004 April 2.5, assuming that the change in the brightness is negligible at the late phase.
Wavelength was calibrated using the standard pixel-wavelength relation of CISCO, 
of which the systematic error is estimated to be less than 0.5 pixels ($<3$\AA).

\subsection{SN 2003hv}
SN 2003hv was discovered by LOTOSS on 2003 Sep. 9.5 (UT) in NGC 1201 at about 12.5mag \citep{beutler03},
and confirmed to be a SN Ia by the spectrum taken on 2003 Sep. 10.4 
which resembles that of SN 1994D two days after maximum \citep{dressler03}.
We therefore assume maximum light in the $B$-band to be JD2452891.

Our $JH$-band spectroscopy of SN 2003hv was carried out 
using OHS/CISCO on 2004 Oct 6 (epoch $+394^{\rm d}$).
The $JH$-band spectroscopy consists of 6 frames of 2000 sec exposure with an 0\farcs5 slit, providing 
a wavelength resolution of $\sim 400$. 
The A2 star SAO 169939 was observed after the target to correct the atmospheric and instrumental 
absorption pattern.
The data were reduced in the same manner as that of SN 2003du.

\subsection{SN2005W}
SN 2005W was discovered on 2005 Feb. 1.4 (UT) in NGC 691 at about 15.2mag \citep{nakano05}, 
and confirmed to be a SN Ia about a week before maximum on 2005 Feb 2.7 (UT) \citep{elias05}.
The expansion velocity is measured to be $\sim 11600$ km s$^{-1}$ from the Si {\sc ii} line.
We assume maximum light in the $B$-band to be 2005 Feb 10 (JD 2453413).

Our $JH$-band spectroscopy of SN 2005W
was carried out on 2005 Sep 12 (epoch $+214^{\rm d}$) by OHS/CISCO.
It consists of 4 frames of 1000 sec exposure with an $0\farcs5$ slit. 
The A2 star HIP 20091 was observed after the target to correct the atmospheric and instrumental 
absorption pattern.
The data were reduced in the same manner as that of SN~2003du.

\section{Results}
The NIR spectra are shown in Figure \ref{specall}.
It can be seen that all the observed 
SNe Ia exhibit strong \feii\ $1.257\ \micron$ and $1.644\ \micron$ lines. 
We discovered two important features. 

First, the line center of \feii , corrected 
for the redshift of the host galaxies, 
is not identical for the three events.
With respect to the rest wavelength of the line ($1.644\ \micron$), 
2 SNe Ia (2003du, 2003hv) 
show a blueshift, while the other SN (2005W) shows a redshift. 
The corresponding velocity shifts relative to the hosts' rest frame
are $-2600$ km s$^{-1}$, $-2100$ km s$^{-1}$, and
$+1400$ km s$^{-1}$, for SNe 2003hv, 2003du, and 2005W, 
respectively.

Such a velocity shift is confirmed by 
a mid-infrared(MIR) spectrum of SN 2003hv 
taken by the IRS/Spitzer Space Telescope at $\sim +360^{\rm d}$ 
after maximum brightness (Fig. \ref{modelfithv}; Gerardy et al., in prep.). 
One of the strongest emission features is identified as the 
ground-state fine-structure line of \coiii\ at $11.89\ \micron$ (a$^{4}$F$_{9/2}$ - a$^{4}$F$_{7/2}$).
The well-isolated \coiii\ line also shows a velocity shift which is consistent with that 
seen in the NIR \feii\ $1.644\ \micron$ line, to within the noise level of the MIR feature. 

Secondly, we find that the observed \feii\ lines show a large variety in their shape
(Figure\ \ref{specall}). 
The spectrum of SN 2003hv clearly shows a flat-topped
boxy profile like that seen at much lower $S/N$ in SN 2003du.
Indeed, the higher $S/N$ of the SN 2003hv spectrum places a
much stronger constraint on the flatness of the core of the 
\feii\ line.
In contrast, the \feii\ line profile of 
SN 2005W shows no evidence of a central flat top.

The boxy profile of \feii\ lines seen in SN 2003du and SN 2003hv 
suggests the absence of radioactive 
\Nifs\ below $\sim 3000$ km s$^{-1}$ (measured from the 
center of the \Nifs\ distribution). 
We note that the observed epoch of SN 2003hv is more advanced 
than SN 2003du. 
At a later epoch, the line profile should follow 
the \Nifs\ distribution more closely (\S 4).
Effects of possible line deformation due to electron and 
line scatterings will also effectively vanish. 
The present result of SN 2003hv confirms the 
existence of the \Nifs-empty hole more strongly than that of SN 2003du.

\section{Discussion}
In this section we discuss the impact of our findings 
for understanding the explosion mechanism. 
It is widely agreed that the explosion of a SN Ia starts  
from a deflagration \citep{nomoto76}.
In the spherical deflagration models, electron capture leads to 
the synthesis of $^{58}$Ni, $^{54}$Fe, and $^{56}$Fe (not via $^{56}$Ni decay), 
thus creating an almost $^{56}$Ni-empty hole \citep[e.g.,][]{nomoto84}.
In multi-D models, the ignition of the deflagration may be off-center, 
producing a non-spherical burning region \citep{wunsch04,plewa04}.
In some cases, 
the deflagration to detonation transition (DDT) may occur 
\citep[delayed detonation models:][]{khokhlov91}.
The DDT may also take place non-spherically 
\citep{livne99} even if the deflagration does not have 
bulk kinematical offset. 

\begin{figure}
\plotone{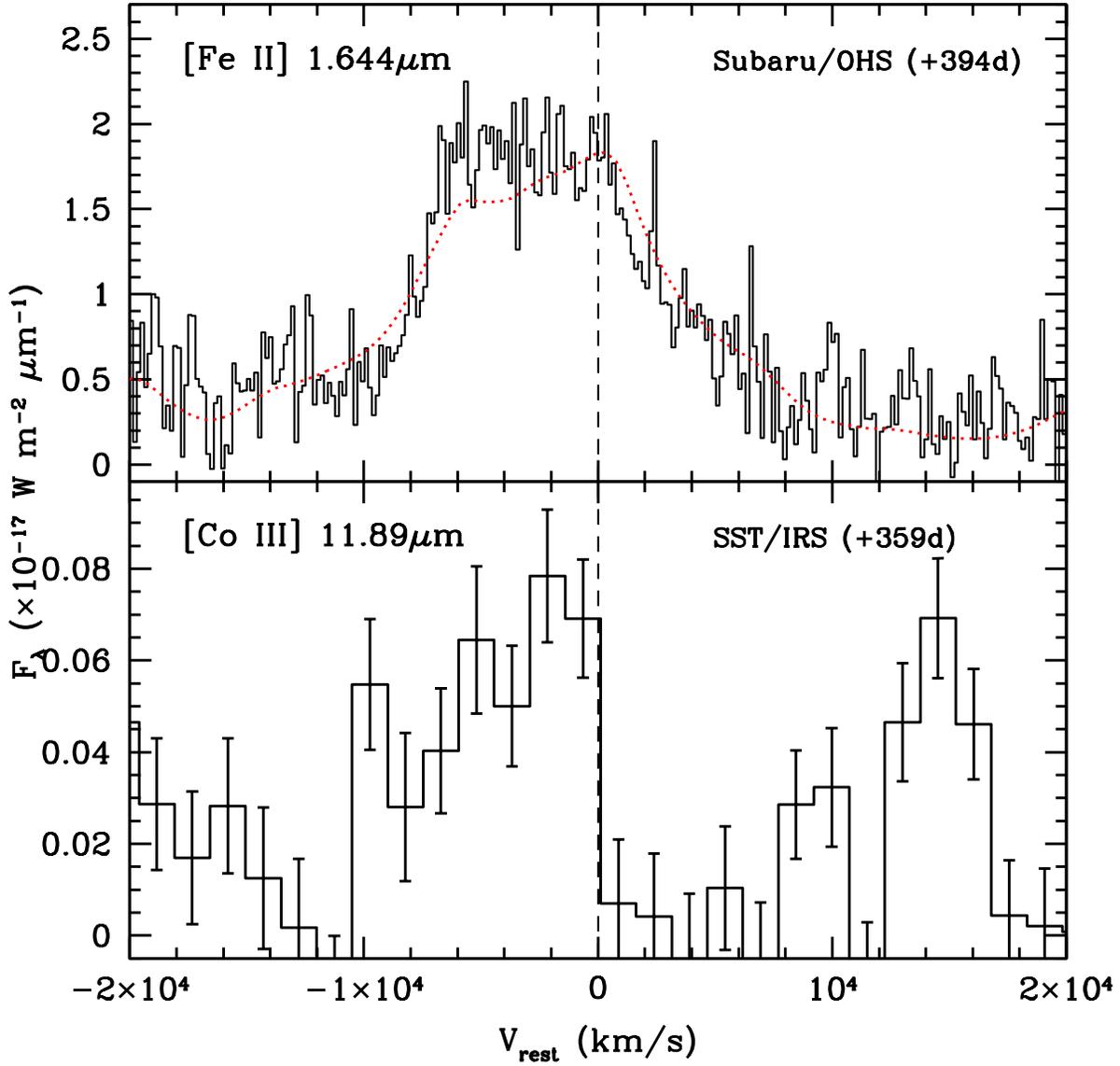}
\caption{\feii\ $1.644\ \micron$ (top) and \coiii\ $11.89\ \micron$ (bottom)  emission lines of 
SN 2003hv (solid).
Overlapped with the \feii\ line is the 1D explosion model W7 at 400$^{\rm d}$
since the explosion (dotted).
The model flux is arbitrarily scaled to fit the observed flux, as well as 
shifted to the blue by 2600 km s$^{-1}$.
}
\label{modelfithv}
\end{figure}

The flat-topped \feii\ NIR lines in SNe 2003hv and 2003du indicate that 
the highest density region occupied by neutron-rich Fe-peak isotopes is not 
mixed with the surrounding region where the dominant isotope is \Nifs\ \citep{hoeflich04}. 
Figure  \ref{modelfithv} shows a spectrum 
for the spherical deflagration explosion model 
W7 (Nomoto et al. 1984) at $400^{\rm d}$ since the explosion, 
compared with that of SN 2003hv. This is calculated by solving transport of $\gamma$-rays 
produced in the decay chain \Nifs\ $\to$ \Cofs\ $\to$ \Fefs\ 
and iteratively solving NLTE rate equations \citep{mazzali01,maeda06}. 
Positrons produced by the \Cofs-decay are assumed to be 
trapped on the spot, since the mean free path of positrons is 
expected to be small at this epoch \citep{milne01}.

The model shows that the flat-topped profile is consistent with the \feii\ $1.644\ \micron$ 
emission for the reason as described above. 
The asymmetry in the profile, mildly peaking in the red, 
is due to the weak contributions of \feii\ $1.664$, $1.677\ \micron$.   

The shift of \feii\ lines suggests that the distribution of 
\Nifs\ (which decays to \Fefs\ via \Cofs)
produced at the explosion is asymmetric, showing 
the bulk kinematical offset of $\gsim 2000$ km s$^{-1}$ 
with respect to the SN rest frame. 
This suggests that, unlike the spherical models, 
the carbon ignition may take place off-center.
The flat-topped \feii\ line profiles suggest 
that electron capture in the high density 
off-center deflagration region creates the neutron-rich hole 
as in 1D models, at least in SNe 2003du and 2003hv. 
The neutron-rich hole is offset along with the bulk \Nifs\
distribution. This indicates that the highest density burning 
in these SNe took place quite far away from the center of the
progenitor star.

 The blueshifted and flat-topped profile of \feii\ can also be reproduced by
 symmetric and opaque dusty core, like \oi\ $\lambda\lambda6300,6364$
 observed in late phase Type II SN \citep{elmhamdi03}.
 If this is the case in the present results,
 MIR \coiii\ line, which will be far less affected by dust extinction,
 may show much smaller blueshift than NIR \feii.
 However, the \coiii\ line shows almost same blueshift as \feii\ (Fig. \ref{modelfithv}).
 Therefore, it is unlikely that the flat-topped profile is caused by the
 opaque dust core.

\begin{figure}
\plotone{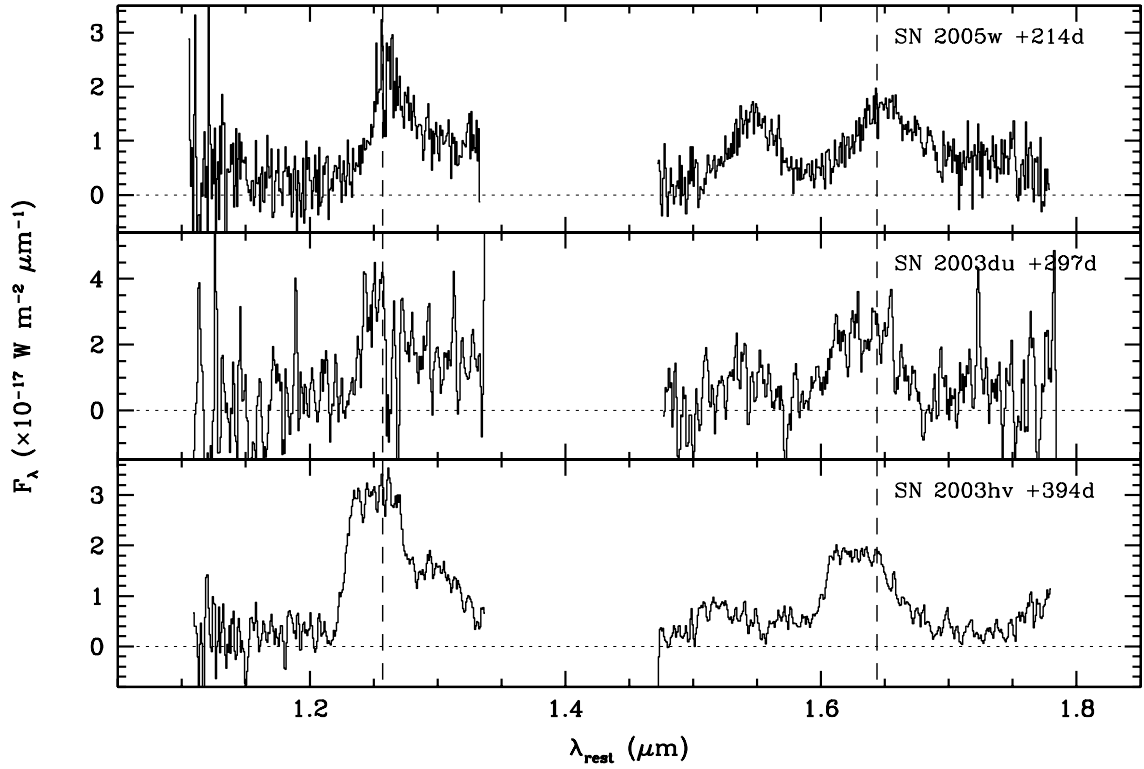}
\caption{NIR $1-2\ \micron$ spectra of late-phase SNe Ia, converted to the rest 
wavelength at the host galaxies.
All the spectra are smoothed with 3 pixel boxcar filter.
Vertical dashed lines show the position of \feii\ $1.257\ \micron$ and $1.644\ \micron$ lines.}
\label{specall}
\end{figure}

Adding the result of SN 1991T \citep{spyromilio92, bowers97}, 
2 SNe (SN 2003du, SN 2003hv) out of 4 SNe Ia show the boxy \feii\ profile (Figure \ref{specall}), 
while SN1991T and SN 2005W do not.
This could indicate that 
the distribution of $^{56}$Ni in the innermost region of the SN Ia explosion might differ  
from object to object. 
SNe Ia with a non-boxy line profile might experience more mixing than 
those with a boxy profile (e.g., SN 2003hv) in the innermost region. 
Another possibility is that the density of the progenitor where 
the ignition takes place differs from object to object, and 
SN Ia with the peaked \feii\ explodes at a low density so as not 
to produce electron capture isotopes.

Alternatively, the different line shape could be due to an age effect, 
i.e., due to a variation in the extent to which the energy deposition from the 
radioactivity is kept local. 
At earlier epochs, $\gamma$-rays are the dominant heating source. 
Since the $\gamma$-ray penetration is not a local process, 
the innermost region, even without \Nifs\ there, can be heated effectively. 
At later epochs, contribution from the local positron energy input becomes larger, 
and the line profile should follow the \Nifs\ distribution more closely. 
Therefore, we expect that the line shape could evolve from a peaked to a 
flat-topped profile.
This may partially explain 
the fact that the most aged SN Ia 2003hv shows the clean flat-topped boxy 
\feii\ line. 
Indeed, this provides an observational test of our model. 

We suggest a future test to estimate the age effect and confirm 
our interpretation of the $^{56}$Ni: 
a temporal series of NIR nebular spectra for an individual SN Ia.
The shape of \feii\ line is expected to start getting flat at
$\sim 250^{\rm d}$, and may 
become totally flat at $\sim 500^{\rm d}$.
Thereafter, it may start showing a peaked profile 
at $\sim 1,000^{\rm d}$ again, 
depending on the amount of positrons that can 
escape the ejecta \citep{milne01}.
Also, spectra at different wavelengths from optical 
to the MIR will be useful to investigate the distribution of different 
ions and heating radioactive isotopes (e.g., Figure  \ref{modelfithv}). 
Because the line profiles from different 
ions and from different energy levels are dependent on the ionization and thermal 
structure of the ejecta, further detailed theoretical study is necessary 
to make use of these observations efficiently.  

\acknowledgments
We thank all the staff at the Subaru observatory 
for their excellent support of the observations.
This work has been supported in part by the Grant-in-Aid for
Scientific Research (17030005, 17033002, 18104003, 18540231) and the
21st Century COE Program (QUEST) from the JSPS and MEXT of Japan.
CLG is supported through UK PPARC grant PPA/G/S/2003/00040.
JCW is supported by NSF Grant AST-0406740.

\clearpage

\begin{deluxetable}{ccclcrrcc}
\tabletypesize{\scriptsize}
\tablecaption{Infrared spectroscopy observing log}
\tablewidth{0pt}
\tablehead{
\colhead{Name} & \colhead{Host} & \colhead{$z_{\rm Host}$} &
\colhead{Obs. Date (UT)}&  \colhead{Epoch\tablenotemark{a} (d)} & \colhead{$\lambda\lambda$(\AA)} & 
\colhead{$\lambda$/$\Delta\lambda$} & \colhead{Exposure (s)} & \colhead{Slit ($^{\prime\prime}$)} 
}
\startdata
2003du &UGC 9391&0.006384& 2004 Feb 27.5  & +297.2 & 11080--18040 & 400 & 4000 & 0.5  \\
2003hv &NGC 1201&0.005604& 2004 Oct 06.5  & +394 & 11080--18040 & 400 & 12000 & 0.5  \\
2005W &NGC 691& 0.008889& 2005 Sep 12.6  & +214 & 11080--18040 & 400 & 4000 & 0.5 
\enddata
\tablenotetext{a}{Epochs are with respect to the epoch of $B$-band
maximum light ($t_{B \rm max}=0$).}
\end{deluxetable}

\clearpage
\onecolumn

\clearpage

\clearpage

\begin{thebibliography}{}
\bibitem[Anupama, Sahu, \& Jose(2005)]{anupama05} Anupama, G. C., Sahu, D. K., \& Jose, J. 2005, A\&A, 429, 667
\bibitem[Beutler \& Li(2003)]{beutler03} Beutler, B. \& Li, W. 2003,  \iaucirc  8197
\bibitem[Bowers et al.(1997)]{bowers97} Bowers, E. J. C., et al. 1997, MNRAS, 290, 663 
\bibitem[Dressler et al.(2003)]{dressler03} Dressler, A., Phillips, M., Morrell, N., \& Hamuy, M. 2003,  \iaucirc, 8198
\bibitem[Elias-Rosa et al.(2005)]{elias05} Elias-Rosa, N., Navasardyan, H., Harutunyan, A., Benetti, S., Turatto, M., Pastorello, A., \& Patat, F. 2005,  \iaucirc, 8479
\bibitem[Elmhamdi et al.(2003)]{elmhamdi03} Elmhamdi, A., et al. 2003, MNRAS, 338, 939
\bibitem[Gamezo et al.(2003)]{gamezo03} Gamezo, V. N., Khokhlov, A. M., Oran, E. S., Chtchelkanova, A. Y., \& Rosenberg, R. O. 2003, Science, 299, 77
\bibitem[Hillebrandt \& Niemeyer(2000)]{hillebrandt00} Hillebrandt, W., \& Niemeyer, J. C. 2000, ARA\&A, 38, 191
\bibitem[H\"oflich et al.(2004)]{hoeflich04} H\"oflich, P., Gerardy, C. L., Nomoto, K., Motohara, K., Fesen, R. A., Maeda, K., Ohkubo, T., \& Tominaga, N. 2004,  \apj, 617, 1258 
\bibitem[Iwamuro et al.(2001)]{iwamuro01} Iwamuro, F., Motohara, K., Maihara, T., Hata, R., \& Harashima, T. 2001,  \pasj, 53,  335 
\bibitem[Khokhlov(1991)]{khokhlov91} Khokhlov, A.M. 1991, A\&A, 245, 114
\bibitem[Kotak et al.(2003)]{kotak03} Kotak, R., Meikle, W. P. S., \& Rodriuez-Gil, P. 2003,  \iaucirc,  8122
\bibitem[Livne(1999)]{livne99} Livne, E. 1999, \apj, 527, L97
\bibitem[Maeda et al.(2006)]{maeda06} Maeda, K., Nomoto, K., Mazzali, P.A., \& Deng, J. 2006, \apj, 640, 854 
\bibitem[Mazzali et al.(2001)]{mazzali01} Mazzali, P. A., Nomoto, K., Patat, F., \& Maeda, K. 2001, \apj, 559, 1047
\bibitem[Milne et al.(2001)]{milne01} Milne, P. A., The, L. -S., \& Leising, M. D. 2001, \apj, 559, 1019
\bibitem[Motohara et al.(2002)]{motohara02} Motohara, K., et al., T. 2002, \pasj, 54, 315 
\bibitem[Nakano \& Li(2005)]{nakano05} Nakano, S. \& Li, W. 2005,  \iaucirc,  8475
\bibitem[Nomoto, Sugimoto, \& Neo(1976)]{nomoto76} Nomoto, K., Sugimoto, D., \& Neo, S. 1976, Ap\&SS, 39, L37 
\bibitem[Nomoto, Thielemann, \& Yokoi(1984)]{nomoto84} Nomoto, K., Thielemann, F. -K., \& Yokoi, K. 1984, \apj, 286, 644 
\bibitem[Nomoto et al.(1994)]{nomoto94} Nomoto, K., Yamaoka, H., Shigeyama, T., Kumagai, S., \& Tsujimoto, T. 1994, in Supernovae, Les Houches Session LIV, ed. S.A. Bludman et al. (Amsterdam: North-Holland), 199
\bibitem[Perlmutter et al.(1999)]{perlmutter99} Perlmutter, S., et al. 1999, ApJ, 517, 565
\bibitem[Piersanti et al.(2003)]{piersanti03} Piersanti, L., Gagliardi, S., Iben, I., Jr., \& Tornamb\'e, A. 2003a, ApJ, 583, 885
\bibitem[Plewa, Calder, \& Lamb(2004)]{plewa04} Plewa, T., Calder, A. C., \& Lamb, D. Q. 2004, \apj, 612, L37
\bibitem[Riess et al.(1998)]{riess98} Riess, A. G., et al. 1998, \aj, 116, 1009 
\bibitem[R\"opke et al.(2006)]{ropke06} R\"opke, F. K., Gieseler, M., Reinecke, M., Travagio, C., \& Hillebrandt, W. 2006, A\&A, 453, 203 
\bibitem[Saio \& Nomoto(2004)]{saio04} Saio, H., \& Nomoto, K. 2004, ApJ, 615, 444
\bibitem[Schwartz \& Holvorcem(2003)]{schwartz03} Schwartz, M., \& Holvorcem, P. R. 2003,  \iaucirc,  8121
\bibitem[Spyromilio et al.(1992)]{spyromilio92} Spyromilio, J., Meikle, W. P. S., Allen, D. A., \& Graham, J. R. 1992, MNRAS, 258, 53
\bibitem[Spyromilio et al.(2004)]{spyromilio04} Spyromilio, J., Gilmozzi, R., Sollerman, J., Leibundgut, B., Fransson, C., \& Cuby, J-G. 2004, A\&A, 426, 547 
\bibitem[Uenishi, Nomoto \& Hachisu(2003)]{uenishi03} Uenishi, T., Nomoto, K., \& Hachisu, I. 2003, \apj, 595, 1094
\bibitem[Wang et al.(2003)]{wang03} Wang, L., et al. 2003, \apj, 591, 1110
\bibitem[Wunsch \& Woosley(2004)]{wunsch04} Wunsch, S., \& Woosley, S. E. 2004, \apj, 616, 1102
\bibitem[Yoon \& Langer(2003)]{yoon03} Yoon, S. -C., \& Langer, N. 2003, A\&A, 312, L53
\end{thebibliography}
\end{document}